\documentstyle[prl,aps,epsf]{revtex}
\input epsf

\begin{document}
\draft

\title{Elementary Derivative Tasks and Neural Net Multiscale Analysis of Tasks}

\author{B.G. Giraud}
\address{giraud@spht.saclay.cea.fr, Service de Physique Th\'eorique, DSM, 
CE Saclay, F-91191 Gif/Yvette, France}

\author{and}

\author{A. Touzeau}
\address{Ecole Centrale de Lyon, 36 avenue Guy de Colongues, 69130 Ecully, 
France}

\date{\today}
\maketitle

\begin{abstract}

Neural nets are known to be universal approximators. In particular, formal 
neurons implementing wavelets have been shown to build nets able to 
approximate any multidimensional task. Such very specialized 
formal neurons may be, however, difficult to obtain biologically and/or 
industrially.  In this paper we relax the constraint of a strict ``Fourier 
analysis'' of tasks. Rather, we use a finite number of more realistic formal 
neurons implementing elementary tasks such as ``window'' or ``Mexican hat'' 
responses, with adjustable widths. This is shown to provide a reasonably 
efficient, practical and robust, multifrequency analysis. A training 
algorithm, optimizing the task with respect to the widths of the responses, 
reveals two distinct training modes. The first mode induces some of the formal 
neurons to become identical, hence promotes ``derivative tasks''. The other 
mode keeps the formal neurons distinct.

\end{abstract}

\section{Introduction}

\medskip
The ability of neural nets to be universal approximators has been proved by 
\cite{Cyben,Hornik} and studied by further authors in different 
contexts. For instance, neurons or small neuronal groups implementing 
``plane wave responses'' have been considered by \cite{Irie} and \cite{Gir1}. 
As well, pairs of neurons implementing ``windows'' have been investigated by 
\cite{Gir2}. Any ``complete enough'' basis of functions which is able to 
span a sufficiently large vector space of response functions is of interest, 
and, for instance, the wavelet analysis has been the subject of a complete 
investigation by \cite{Benve} and \cite{Drey}.

\medskip
In this paper, we visit again the subject of a linear reconstruction of 
tasks, but with an emphasis upon neglecting the usual ``translational'' 
parameters. We mainly use a scale parameter only. This is somewhat 
different from the usual wavelet approach, which takes advantage of both 
translation and scale. But we shall find that a multifrequency reconstruction 
of tasks occurs as well. Simultaneously, we separate a ``radial'' from an 
``angular'' analysis of the task. Finally, for the sake of robustness and 
biological relevance, we introduce a significant amount of randomness, 
corrected by training, in the choice of the implemented neuronal parameters. 
Furthermore, our basic neuronal units can be those ``window-like'' pairs 
advocated earlier \cite{Gir2}, because of biological relevance too. Such 
deviations from the more rigorous approaches of \cite{Benve} and \cite{Drey} 
are expected to make cheaper the practical implementation of such neural nets.

\medskip
We also investigate two training operations. The first one consists in a 
trivial optimization of the output synaptic layer connecting a layer of 
intermediate, ``elementary task neurons'' to an output, purely {\it linear} 
neuron. The second training consists in optimizing the scale parameters of 
such a layer of intermediate neurons. It will be found that one may start from 
random values of such parameters and, however, sometimes reach solutions where 
some among the intermediate neurons are driven to become identical. This 
``dynamical identification'' training will be discussed.

\medskip
In Section II we describe our formalism, including a traditional universality 
theorem. We also reduce the realistic, multi-dimensional situations to a 
one-dimensional problem. In Section III we illustrate such considerations by 
numerical examples of network training. Finally Section IV contains our 
discussion and conclusion.

\section{Formalism}

\subsection{Definitions, architecture}

\medskip
Consider an input $X>0$ which must be processed into an output (a task)
$F(X).$ This input is here taken to be a positive number, such as the 
intensity of a spike or the average intensity (or frequency) of a spike 
train. One may view $X$ as a ``radial'' coordinate in a suitable space. 
There is no loss of generality in restricting $X$ to be a positive number, 
because, should negative values of $X$ be necessary for the argument, then $F$ 
could always be split into an even and odd parts, 
$\left[F(X)\pm F(-X)\right]/2,$ respectively. Such even and odd parts need 
only be known for $X>0,$ obviously. Outputs, in turn, will have both signs, 
in order to account for both excitation or inhibition. Finally there is no 
need to tell a scalar task $F(X)$ from a vector task $\{F_1,F_2,...\},$ since 
any component $F_k$ boils down to a separate scalar task, and this can be 
processed by a parallel architecture.

\medskip
Consider now neuronal units which, for instance may be excitatory-inhibitory 
pairs of neurons providing a window-like elementary response. Or they may be
more complicated assemblies of neurons, providing a more elaborate ``mother 
wavelet'', such as a ``Mexican hat''. We denote  $f(X)$ the response 
function of such a unit and, for short, call this unit a ``formal neuron'' 
(FN). The traditional wavelet approach uses a set of such FN's
with various thresholds $b$ and scale sensitivities $\lambda,$ hence a 
space of elementary responses $f\left[(X-b)/\lambda\right].$ The same approach 
expands $F$ in this set,
\begin{equation}
F(X)=  \int db \, d\lambda \  w(b,\lambda) \  f\left[(X-b)/\lambda)\right],
\end{equation}
where the integral is most often reduced to a discrete sum. Also,
$b$ and  $\lambda$  do not need to be independent parameters. The expansion 
coefficients, $w(b,\lambda),$ are output synaptic weights and are the 
unknowns of the problem. This well known architecture is shown in Figure 1.

\begin{figure}[htb] \centering
\mbox{  \epsfysize=80mm
         \epsffile{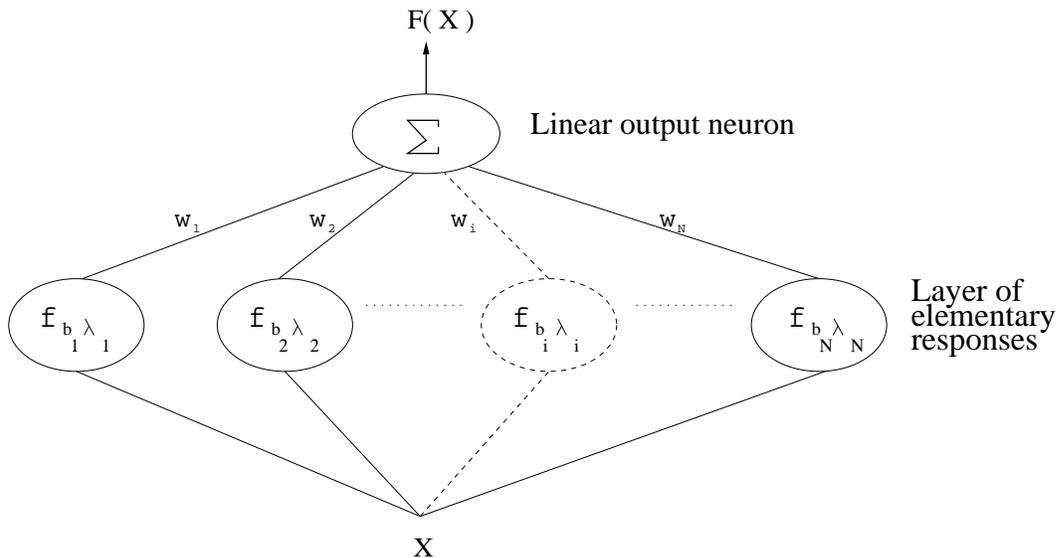}
     }
\caption{All elementary units (FN's) receive the same input $X$. Each unit 
returns an output $f,$ which depends on parameters such as a threshold $b$
and a gain $\lambda.$ Output synaptic weights $w(b,\lambda)$ linearly mix such 
elementary outputs $f(X;b,\lambda)$ into a global output $F(X).$}
\end{figure}

\subsection{One-dimensional universality, radial case}

\medskip
The following, seemingly poorer, but simpler expansion,
\begin{equation}
F(X)= \int_0^\infty d\lambda \  w(\lambda) \ \lambda^{-1} f\left(X/\lambda
\right),
\label{basicinte}
\end{equation}
does not use the translation parameter $b.$ Here it is assumed that there 
exists a suitable electronic or biological tuning mechanism, able to recruit 
or adjust FN's with suitable gains $\lambda^{-1},$ but no threshold tuning.
Such gains are positive numbers, naturally. The outputs of such FN's are 
then added, via synaptic output efficiencies $w(\lambda),$ which can be 
both positive and negative, namely excitatory and inhibitory, respectively. 
The coefficient $\lambda^{-1}$ is introduced in Eq. (\ref{basicinte})
for convenience only. It can be absorbed in $w(\lambda).$

\medskip
This expansion, Eq. (\ref{basicinte}) allows a universality theorem. Define 
$Y=\ln X$ and $L=\ln \lambda.$ The same expansion becomes,
\begin{equation}
G(Y) \equiv F\left(e^Y\right)= \int_{-\infty}^\infty dL \  W(L) \ g(Y-L),
\end{equation}
where $W(L) \equiv w\left(e^L\right)$ and 
$g(Y) \equiv f\left(e^Y\right).$ This reduces the ``scale expansion'',
Eq. (\ref{basicinte}), into a ``translational expansion'' where a basis
is generated by arbitrary translations of a given function. The solution of 
this inverse convolution problem  is trivially known as 
${\hat W}(p)={\hat G}(p)/{\hat g}(p),$ where the superscript 
${\hat{ }}$ refers to the Fourier transforms of $ W,$ $G$ and $g,$ 
respectively, and $p$ is the relevant ``momentum''. This result will
make our claim for universality. In the following, this paper empirically
assumes that the needed analytical properties of $f$, $\hat f$, ...$\hat W$ 
are satisfied. Actually, for the sake of biological or industrial relevance, 
we are only concerned with discretizations of Eq. (\ref{basicinte}), with
$N$ units,
\begin{equation}
F_{app}(X) = \sum_{i=1}^N  w(\lambda_i) \ f\left(X/\lambda_i\right),
\label{pratiq}
\end{equation}
where we now let $w$ include the coefficient $\lambda_i^{-1}\, d\lambda.$

\subsection{Rotational analysis}

\medskip
Obviously, input patterns to be processed by a net cannot be reduced to one 
degree of freedom $X$ only. Rather, they consist of a vector $\vec X$ 
with many components $X_1,X_2,...,X_P.$ These may be considered as, and 
recoded into, a radial variable $X=\sqrt{\sum_{j=1}^P X_j^2}$ and, to specify 
a direction on the suitable hypersphere, $(P-1)$ angles $\alpha_1, \alpha_2,
 ...,\alpha_{P-1}.$ Enough special functions (Legendre polynomials, spherical 
harmonics, rotation matrices, etc.) are available to generate complete 
functional bases in angular space and one might invoke some formal neurons 
as implementing such base angular functions. The design of such FN's, and as 
well the design of such a polar coordinate recoding, is a little far fetched, 
though. In this paper we prefer to take advantage of the following argument, 
based upon the synaptic weights of the input layer, shown in Figure 2.
\begin{figure}[htb] \centering
\mbox{  \epsfysize=80mm
         \epsffile{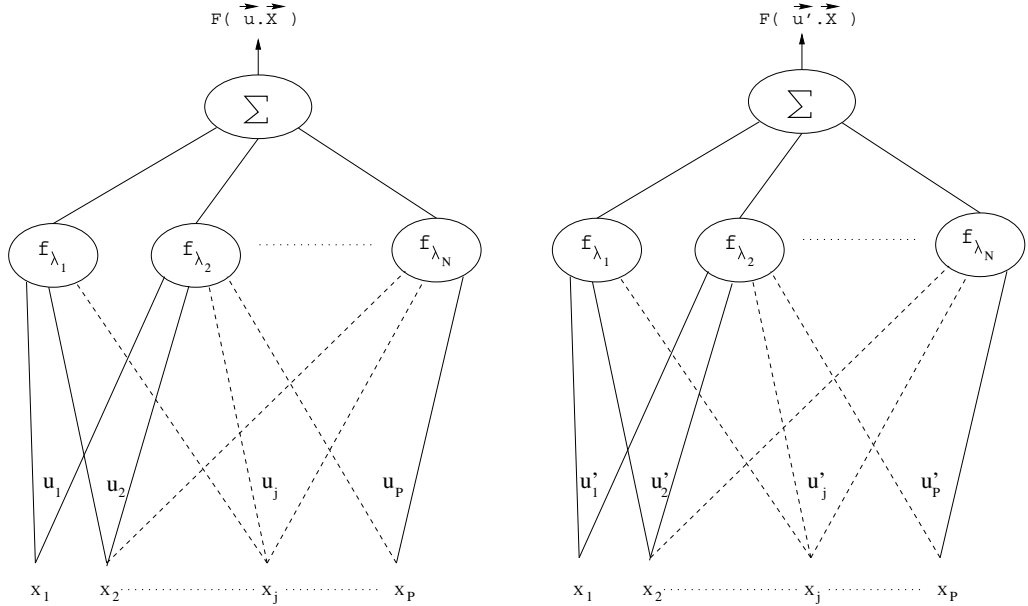}
     }
\caption{Architecture showing how a task can be rotated by means of the input
synaptic weights.}
\end{figure}
In the left part of the Figure, Fig. 2, all the FN's have the same input
synaptic weights $\vec u \equiv \{u_1,u_2,...,u_P\},$ hence receive
the same input $X=\vec u \cdot \vec X$ when contributing to a global task $F.$
For the right part of Fig. 2 it is again assumed that all FN's have equal 
input weights, with, however, weights $\vec u'$ deduced from $\vec u$ by a 
sheer rotation, $\vec u' = {\cal R} \vec u.$  Accordingly, if the output 
weights of the left part are the same as those of the right one, the global 
task $F(\vec u' \cdot \vec X)$ performed by the right part is a rotated task, 
$F'={\cal R} F.$ An expansion of any task ${\cal F}$ upon the 
$(P-1)$-rotation group is thus available,
\begin{equation}
{\cal F}=\int d{\cal R} \ {\cal W}\left({\cal R}\right)\, {\cal R} F,
\end{equation}
where discretizations are in order, naturally, with suitable output weights 
${\cal W}.$ Here $F$ plays the r\^ole of an elementary task, and it might be
of some interest to study cases where $F$ belongs to specific representations 
of the rotation group. This broad subject exceeds the scope of the present 
paper, however, and, in the following, we restrict our considerations to 
scalar tasks $F(X)$ of a scalar input $X,$ according to Fig. 1 only.

\subsection{Training output weights}

\medskip
Let us return to Eq. (\ref{pratiq}), in an obvious, short notation 
$F_{app}=\sum_i w_i f_i.$ Two kinds of parameters can be used to best 
reconstruct $F$: the output synaptic weights $w_i$ and, hidden inside the 
elementary tasks $f_i,$ the scales $\lambda_i.$ Let $\langle \, | \, \rangle$ 
denote a suitable scalar product in the functional space spanned by all the 
$f_i$'s of interest. We assume, naturally, that the same scalar product makes 
sense for the $F$'s. Incidentally, there is no loss of generality if $F$ is 
normalized, $\langle F | F \rangle =1,$ since the final neuron is linear.

\medskip
One way to define the ``best'' $F_{app}$ is to minimize the square norm of the
error $(F-F_{app}).$ In terms of the $w_i$'s, this consists in solving the
equations,
\begin{equation}
\frac{\partial}{\partial w_i}\left( \langle F | F \rangle - 2 
\sum_{j=1}^N w_j \langle f_j | F \rangle + \sum_{j,k=1}^N w_j 
\langle f_j | f_k \rangle w_k \right)=0, \ \ i=1,...,N.
\label{linear}
\end{equation}
Let ${\cal G}$ be that matrix with elements 
${\cal G}_{jk}=\langle f_j | f_k \rangle.$
Its inverse ${\cal G}^{-1}$ usually exists. Even in those rare cases when 
${\cal G}$ is very ill-conditioned, or its rank is lower than $N,$ it is easy
to define a pseudoinverse such that, in all cases, the operator 
${\cal P}=\sum_{i,j=1}^N | f_i \rangle \left({\cal G}^{-1}\right)_{ij} 
\langle f_j |$
is the projector upon the subspace spanned by the $f_i$'s. Then a trivial 
solution, $F_{app}={\cal P}F,$ is found for Eqs. (\ref{linear}),
\begin{equation}
w_i=\sum_{j=1}^N \left({\cal G}^{-1} \right)_{ij} \langle f_j | F \rangle,
\ \ i=1,...,N.
\end{equation}
Given $F$ and the $f_i$'s, this projection, which can be achieved by 
elementary trainings of the output layer of synaptic weights, will be 
understood in the following. It makes the $w_j$'s functions of the 
$\lambda_i$'s.

\subsection{Training elementary tasks}

\medskip
Now we are concerned with the choice of the parameters $\lambda_i$ of
the FN's performing elementary tasks. This is of some importance, for the
number $N$ of FN's in the intermediate layer is quite limited in practice. 
The subspace spanned by the $f_i$'s is thus most undercomplete.
Hence, every time one requests an approximator to a new $F,$ an 
optimization with respect to the intermediate layer is in order, to patch
likely weaknesses of the ``projector'' solution, Eqs (\ref{linear}).

\medskip 
Let us again minimize the square norm 
${\cal E}=\langle \, (F-F_{app}) \, | \, (F-F_{app}) \, \rangle$ of the error. 
We know from Eqs. (\ref{linear}) that the $w_i$'s are functions of the 
$\lambda_j$'s, but there is no need to use chain rules 
$\partial w_i/\partial \lambda_j\ \partial/\partial w_i,$
because the same equations, Eqs. (\ref{linear}), cancel the corresponding 
contributions, the $w_i$'s being optimal. Derivatives of $f_i$ with respect 
to their scales $\lambda_i$ are enough. The gradient of ${\cal E},$ to be 
cancelled, reads,
\begin{equation}
\frac{\partial {\cal E}}{\partial \lambda_j}=\frac{2 w_j}{\lambda_j^2}\,
\langle \, X f'(X/\lambda_j)\, | \, (F-F_{app}) \, \rangle =0, \ \ j=1,...,N.
\label{gradien}
\end{equation}
Here $f'$ is the straight derivative of the reference elementary task, before
any scaling. There is no difficulty in implementing a training algorithm for 
a gradient descent in the $\lambda$-space.

\medskip
The next section, Sec. III, gives a brief sample of the results we obtained 
when solving Eqs. (\ref{linear}) and (\ref{gradien}) for many choices of the 
global task $F$ and elementary task $f.$

\section{Numerical Illustrative Examples}

\subsection{Symmetry and degeneracy}

\medskip
Define for instance the scalar product in the functional space as,
$\langle f_i | f_j \rangle \equiv \int_0^{20} dX\, f_i(X)\, f_j(X),\, \forall 
\, f_i,f_j.$ 
Among many numerical tests we show here the results obtained when the target 
task reads, $F(X)=0.10167\, e^{-X/10}
\{ 0.60717 \tanh[4(X-1.66133)] - 4.33575 \tanh[4(X-9.56591)] \}.$ Let the 
elementary task of a FN read 
$f(X/\lambda)=(1-X^2/\lambda^2)\, e^{-X^2/(2\lambda^2)},$ a Mexican hat. Set 
$N=5,$ and initial values $1/4, 1/2, 1, 2,4$ for the $\lambda_i$'s. Keeping 
Eqs. (\ref{linear}) satisfied at each step, start a gradient descent from such
initial values. Our increments of the $\lambda_i$'s at each step read,
$\delta \lambda_i=-2 \partial {\cal E} / \partial \lambda_i,$ see 
Eqs. (\ref{gradien}). After $\simeq 90$ steps, a saturation of 
$||F_{app}||^2=\langle F | {\cal P} | F \rangle $ begins, see Figure 3.
\begin{figure}[htb] \centering
\mbox{  \epsfysize=80mm
         \epsffile{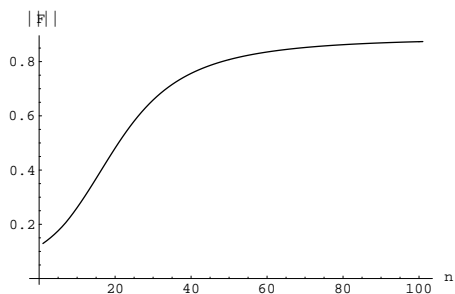}
     }
\caption{Learning curve: the norm $||F||$ of $F_{app}$ increases as 
a function of the number $n$ of learning steps, then saturates.}
\end{figure}
A comparison between $F$ and $F_{app}$ is provided by Figure 4.
\begin{figure}[htb] \centering
\mbox{  \epsfysize=80mm
         \epsffile{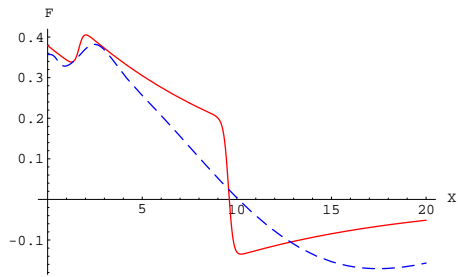}
     }
\caption{A target task (solid line) and its best approximation (dashed) after 
learning saturation.}
\end{figure}
This saturation makes it reasonable to interrupt the learning process. For the 
sake of rigor, however, another run, with 1000 steps, was used to verify the 
saturation. While the saturation is confirmed, the convergence of the 
$\lambda_i$'s turns out to be slower. The values of the $\lambda_i$'s and 
$w_i$'s at the end of this second run read 
$\{0.249,0.535,1.0512,1.0522,11.13\}$ and 
$\{-0.0008,-0.0002,38.107,-38.121,0.3764\},$ respectively. The weakness of 
$w_1$ and $w_2$ is explained by the lack of a fine structure in $F.$ The large
and almost opposite values of $w_3$ and $w_4$ clearly mean a renormalization of
$(f_3-f_4),$ since $\lambda_3$ and $\lambda_4$ are so close to each other. 
Numerical difficulties linked to difference effects may therefore demand extra 
care in practical applications. We show in Figure 5 the way in which the 
$\lambda_i$'s evolved during the first 100 steps of the gradient descent.
\begin{figure}[htb] \centering
\mbox{  \epsfysize=100mm
         \epsffile{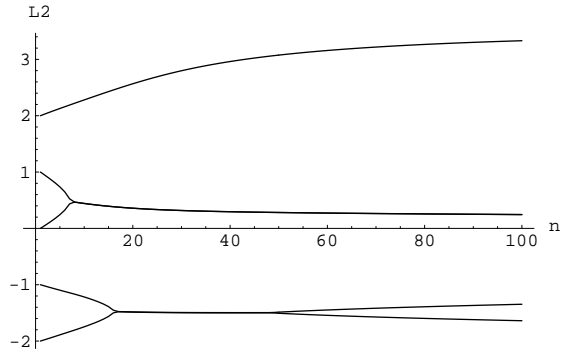}
     }
\caption{Evolution of $Log_2\, \lambda_i,$ $i=1,...5,$ during learning, for
one of the ``derivative sensitivity cases''. Notice the fusions of {\it
two} pairs of scales, then the splitting of one of them.}
\end{figure}
A temporary merging of $\lambda_1$ with $\lambda_2,$ then a final episode 
in which they become distinct again, are striking, as well as the
merging of $\lambda_3$ with $\lambda_4.$ It will be stressed at this stage
that the whole process is invariant under any permutation of the $\lambda_i$'s 
(and of their associated $w_i$'s), hence a ``triangular'' rule, 
$\lambda_i \le \lambda_{i+1}$ can be implemented without restricting learning 
flexibility. Furthermore, as a symmetric function under pairwise exchanges of 
such parameters, the error square norm $\cal E$ has a vanishing ``transverse'' 
derivative, $\partial {\cal E}/\partial (\lambda_i-\lambda_j)=0,$ every time 
$\lambda_i=\lambda_j.$ It is thus not surprising that, at least for part of the
learning process, the learning path rides lines where such parameters merge.

\medskip
When merging occurs, the functional basis seems to degenerate since $f_i$ and
$f_j$ are not distinct. It will be recalled, however, that our output neuron
is linear, and nothing prevents the process from using the strictly equivalent
representations, 
$w_i f_i + w_j f_j \equiv (w_i+w_j)/2 \times (f_i+f_j) + (w_i-w_j)/2 \times
(f_i-f_j).$ A trivial renormalization of the $(f_i-f_j)$ term makes it that 
the functional basis still contains two independent vectors, namely, a new 
elementary response $\partial f/\partial \lambda$ besides $f_i=f_j.$ Naturally,
the renormalization has a numerical cost, since both $w_i$ and $w_j$ must 
diverge. In practice, a minute modification of the ``triangular rule'', which 
becomes, in our runs, $\lambda_{i+1}-\lambda_i \ge 10^{-3},$ is enough to 
smooth our calculations. The conclusion of this merging phenomenon, for those 
$F$'s where it occurs, is of some interest: new neuronal units (new FN's) may 
spontaneously emerge. These are ``derivative sensitive'', and may represent a 
new task $\partial F/\partial \lambda,$ or, if $(p+1)$ parameters merge,
any further derivative $\partial^p f/\partial \lambda^p.$

\subsection{Full Symmetry Breaking}

\medskip
Most choices of $F$ yield distinct values for the $\lambda_i$'s. We show in 
Figure 6 a trivial case. Here $f=1/[1+(X^2/\lambda^2)],$ a window-like 
elementary response, and the target task reads $F= 9/(1+16X^2) + 5/(1+4X^2) + 
2/(1+X^2) -  1/[1+(X^2/4)] - 1/[1+(X^2/16)],$ a sum of such windows. We 
freeze $\lambda_3=1,$ $\lambda_4=2,$ and $\lambda_5=4,$ a symmetry breaking 
situation, and clearly a part of the obvious solution for the minimum of 
${\cal E}.$ Then the contour map of ${\cal E}$ in the 
$\{\lambda_1$,$\lambda_2\}$-space does show the expected minimum for 
$\lambda_1=1/4$ and $\lambda_2=1/2.$ The minimum turns out to be very flat,
hence some robustness is likely for that special case. The learning process 
does reach this ``fully symmetry breaking'' configuration, together with the 
corresponding set of $w_i$'s, namely $\{9,5,2,-1,-1\}.$ Many other, less 
academic cases generate a full symmetry breaking, namely distinct 
$\lambda_i$'s.
\begin{figure}[htb] \centering
\mbox{  \epsfysize=80mm
         \epsffile{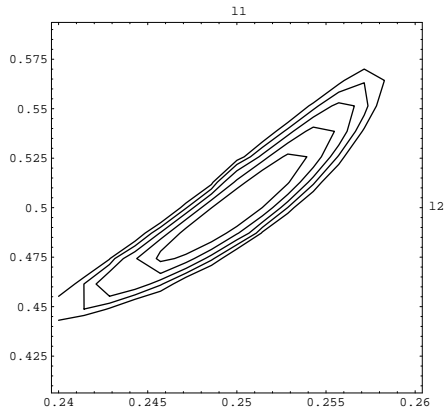}
     }
\caption{Symmetry breaking case. Contours of the error in the vicinity of a 
symmetry breaking set of parameters. The last three out of five adjustable 
parameters are frozen and distinct. The minimum of the error is reached with 
unequal values of the first two parameters.} 
\end{figure}

\subsection{More numerical results}

\medskip
Besides ``windows'' and ``Mexican hats'', we also used oscillatory shapes 
such as $(\sin X)/X$ for $f.$ A cut-off by an exponential decay was also 
sometimes introduced. The range of the scalar product integration was 
independently varied within one order of magnitude. Sometimes the dimension 
$N$ of the elementary task basis was also taken as a random number, a test of 
little interest, however, which just verified that $F_{app}$ improves when $N$ increases. For $F,$ a few among our tests involved a small amount of random 
noise added to a smooth main part $F_{background}.$ Furthermore we 
investigated a fair amount of piecewise continuous $F$'s, this case being of 
interest for image processing \cite{Hertz}. Alternately, we smoothed such 
discontinuities with a suitable definition of $F,$ such as, 
$F=\sum_{\ell} c_{\ell} \tanh[\sigma (X-\beta_{\ell})],$ with randomized 
choices of the number of terms, the coefficients $c_{\ell},$ the large ``slope
coefficient $\sigma,$ and the positions $\beta_{\ell}$ of the steep areas.
The set of initial values for the $\lambda_i$'s before gradient descent was
also sometimes taken at random. It was often found that a traditional sequence
$\lambda_i \simeq 2^{i-(N+1)/2}$ is not a bad choice for a start.

All our runs converge reasonably smoothly to a saturation of the norm 
$||F_{app}||,$ provided those cases where ${\cal G}$ becomes ill-conditioned 
are numerically processed. There is a significant proportion of runs where
the optimum seems to be quite flat, hence some robustness of the results.
Local minima where the learning gets trapped do not seem to occur very often, 
but this problem deserves the usual caution, with the usual annealing if 
necessary. We did not find clear criteria for predicting whether a given $F$
leads to a merging of some $\lambda_i$'s, however. Despite this failure,
all these results advocate a reasonably positive case for the learning
process described by Eqs. (\ref{linear}) and (\ref{gradien}) and the emergence
of ``derivative tasks''.

\section{Discussion and conclusion}

\medskip
This paper tries to relate several issues. Most of them are well known in 
the theory of neural nets, but two of our considerations, the question of 
symmetries and the rotational analysis, might give reasonably original 
results, up to our knowledge at least.

\medskip
The most important and well known issue is that of the universality 
offered by nets whose architecture is described by Figures 1 and 2, 
namely four layers: input weights ${\bf u},$ FN's for elementary tasks 
${\bf f}$ with adjustable parameters ${\bf M},$ output weights ${\bf w},$ 
linear output neuron(s). The linearity of the output(s) can be summarized 
in any dimensions by the linear transform 
${\bf F}({\bf X})=\int d{\bf M}\, {\bf w}({\bf M})\, {\bf f}({\bf X};{\bf M}).$
(We use here boldface symbols to stress that the linearity generalizes to 
any suitable vector and tensor situations for multiparameter inputs, 
intermediate tasks and outputs.)
This linearity actually reduces the theory of such an architecture to a 
special case of the  ``generator coordinate'' theory, well known in physics 
\cite{Hill}. As well, from a mathematical point of view, this boils down to
the only question of the invertibility of the kernel 
${\bf f}({\bf X};{\bf M}).$ 
Actually, the invertibility problem boils down into identifying those 
classes of global tasks ${\bf F}$ which belong to the functional (sub)space 
spanned by the ${\bf f}$'s. For the sake of definiteness, we proved a 
universality theorem for the very special case of ``scaling without 
translating'', inspired by wavelets. But most of the considerations of 
this paper clearly hold if one replaces, {\it mutatis mutandis}, wavelets
by other responses and scaling parameters by any other parameters.

\medskip
The parameters ${\bf M}$ can be defined as including the input synaptic weight 
vectors $\vec u,$ whose dimension is necessarily the same as that of the 
inputs $\vec X$ in order to generate the actual inputs $\vec u \cdot \vec X$ 
received by the intermediate FN's. When ${\bf M}$ also explicitly includes 
scale parameters $\lambda,$ there is no loss of generality in restricting 
the $\vec u$'s to be unitary vectors. Hence the linear kernel ${\bf f}$ 
can imply, in a natural way, an integration upon the group of rotations
transforming all the $\vec u$'s into one another. This part of the theory
relates to the angular momentum projections which are so familiar in the 
theory of molecular and nuclear rotational spectra \cite{Bohr}.

\medskip
The well known issue of the discretization of a continuous expansion converts 
kernels into finite matrices, naturally. This paper studied what happens 
if one trains ${\bf w}$ for a temporary optimum of the approximate task 
${\bf F}_{app}$, while ${\bf M}$ is not yet optimized. This implies a prejudice on 
training speeds: ${\bf w}$ fast learner, ${\bf M}$ slower. Other choices, 
such as ${\bf w}$ slower learner and ${\bf M}$ faster, for instance, are
as legitimate, and should be investigated too. The question is of importance
for biological systems, because of obvious likely differences in the time
behaviors and biochemical and metabolic factors of synapses and cell bodies. 
The training speed hierarchy we chose points to one technical problem only, 
namely whether the Gram-Schmidt matrix ${\cal G}$ of scalar products 
$\langle {\bf f}_i | {\bf f}_j \rangle$ is easily invertible or not. We do 
not use a Gram-Schmidt orthogonalization of the finite basis of such $f_i$'s, 
but the (pseudo) inversion of ${\cal G}$  amounts to the same. Once 
${\cal G}^{-1}$ is obtained, temporarily optimal ${\bf w}$ are easily derived.

\medskip
Our further optimization of ${\bf F}_{app}$ with respect to the parameters
of the intermediate FN's takes advantage of the linearity of the output(s)
and the symmetry of the problem under any permutation of the FN's. Let $i$
label such FN's, $i=1,...,N$ and denote ${\bf M}_i$ the parameters of the 
$i$-th FN. We found cases where the gradient descent used to optimize 
${\bf F}_{app}$ induces a few ${\bf M}_i$'s to become quite close to one 
another. Such functional clusters, because of the output linearity, may yield 
elementary tasks corresponding to derivatives of ${\bf f}$ with respect to 
components of ${\bf M}.$ This derivative process may look similar to a 
Gram-Schmidt orthogonalization, but it is actually distinct, because no rank 
is lost in the basis. For those ${\bf F}$'s which induce such mergings of FN's,
industrial  applications should benefit from a preliminary simulation of 
training as a useful precaution, because, besides straight FN's implementing 
${\bf f},$ additional, more specific $FN's$ implementing ``derivative 
${\bf f}$'s'' will be necessary. For biological systems, diversifications of 
neurons, or groups of such, between tasks and ``derivative tasks'' might also
be concepts of interest. It may be noticed that the word ``derivative'' may 
hold with respect to inputs as well as parameters. Indeed, as found at the 
stage of Eq. (3), scale parameters reduce, in a suitable representation, to
translational parameters in a task $g(Y-L).$ The sign difference between 
$\partial g/\partial Y$ and $\partial g/\partial L$ is obviously 
inconsequential.

\medskip
To conclude, this emergence of ``derivative elementary tasks'' prompts us into
a problem yet unsolved by our numerical studies with many different 
${\bf F}'$s and many different ${\bf f}$'s: given the shape of ${\bf f},$ can 
one predict whether a given ${\bf F}$ leads to a full symmetry breaking or to 
a partial merging of the FN's?

\section{Acknowledgment}

A.T. thanks Service de Physique Th\'eorique, Saclay, for its hospitality 
during this work.

\end{document}